\documentclass[twocolumn,superscriptaddress,prl]{revtex4-1}
\usepackage{amsmath}
\usepackage{amsfonts}
\usepackage{mathrsfs}
\usepackage{graphicx}
\usepackage{times}
\usepackage{color}
\usepackage[normalem]{ulem}
\usepackage{epstopdf}

\newcommand\redout{\bgroup\markoverwith{\textcolor{red}{\rule[.5ex]{2pt}{1pt}}}\ULon}

\def\be{\begin{equation}}
	\def\ee{\end{equation}}
\def\bea{\begin{eqnarray}}
	\def\eea{\end{eqnarray}}
\def\nn{\nonumber}

\begin{document}
	\title{Nonlinearity-induced dynamical self-organized twisted-bilayer lattices in Bose-Einstein condensates}
	
\author{Rui Tian}
\thanks{These authors contribute equally to this work.}
\affiliation{Ministry of Education Key Laboratory for Nonequilibrium Synthesis and Modulation of Condensed Matter,Shaanxi Province Key Laboratory of Quantum Information and Quantum Optoelectronic Devices, School of Physics, Xi'an Jiaotong University, Xi'an 710049, China}

\author{Yue Zhang}
\thanks{These authors contribute equally to this work.}
\affiliation{Ministry of Education Key Laboratory for Nonequilibrium Synthesis and Modulation of Condensed Matter,Shaanxi Province Key Laboratory of Quantum Information and Quantum Optoelectronic Devices, School of Physics, Xi'an Jiaotong University, Xi'an 710049, China}

\author{Tianhao Wu}
\affiliation{Ministry of Education Key Laboratory for Nonequilibrium Synthesis and Modulation of Condensed Matter,Shaanxi Province Key Laboratory of Quantum Information and Quantum Optoelectronic Devices, School of Physics, Xi'an Jiaotong University, Xi'an 710049, China}

\author{Min Liu}
\affiliation{Ministry of Education Key Laboratory for Nonequilibrium Synthesis and Modulation of Condensed Matter,Shaanxi Province Key Laboratory of Quantum Information and Quantum Optoelectronic Devices, School of Physics, Xi'an Jiaotong University, Xi'an 710049, China}

\author{Yong-Chang Zhang}
\affiliation{Ministry of Education Key Laboratory for Nonequilibrium Synthesis and Modulation of Condensed Matter,Shaanxi Province Key Laboratory of Quantum Information and Quantum Optoelectronic Devices, School of Physics, Xi'an Jiaotong University, Xi'an 710049, China}

\author{Shuai Li}
\email{lishuai0999@stu.xjtu.edu.cn}
\affiliation{Ministry of Education Key Laboratory for Nonequilibrium Synthesis and Modulation of Condensed Matter,Shaanxi Province Key Laboratory of Quantum Information and Quantum Optoelectronic Devices, School of Physics, Xi'an Jiaotong University, Xi'an 710049, China}

\author{Bo Liu}
\email{liubophy@gmail.com}
\affiliation{Ministry of Education Key Laboratory for Nonequilibrium Synthesis and Modulation of Condensed Matter,Shaanxi Province Key Laboratory of Quantum Information and Quantum Optoelectronic Devices, School of Physics, Xi'an Jiaotong University, Xi'an 710049, China}
	
\begin{abstract}
Creating crystal bilayers twisted with respect to each other would lead to large periodic supercell structures, which can support a wide range of novel electron correlated phenomena, where the full understanding is still under debate.
Here, we propose a new scheme to realize a nonlinearity-induced dynamical self-organized twisted-bilayer lattice in an atomic Bose-Einstein condensate (BEC).
The key idea here is to utilize the nonlinear effect from the intrinsic atomic interactions to couple different layers and induce a dynamical self-organized supercell structure, dramatically distinct from the conventional wisdom to achieve the static twisted-bilayer lattices. To illustrate that, we study the dynamics of a two-component BEC and show that the nonlinear interaction effect naturally emerged in the Gross-Pitaevskii equation of interacting bosonic ultracold atoms can dynamically induce both periodic (commensurable) and aperiodic (incommensurable) moir\'{e} structures. One of the interesting moir\'{e} phenomena, i.e., the flat-band physics, is shown through investigating the dynamics of the wave packet of BEC. Our proposal can be implemented using available state-of-the-art experimental techniques and reveal a profound connection between the nonlinearity and twistronics in cold atom quantum simulators.
\end{abstract}
\maketitle
	
\section{Introduction}
Superimposed crystal bilayers twisted with respect to each other provide a new knob to unveil a wide
range of novel electron correlated phenomena \cite{dos2007graphene,sboychakov2015electronic,dai2016twisted,liu2020tunable, cao2020tunable, chen2019signatures, shen2020correlated, regan2020mott, tang2020simulation}.
In particular, moir\'{e} patterns associated  with small twist angles result in various novel physical properties \cite{lu2013twisting,woods2014commensurate,yan2013strain, alden2013strain, kim2016van, yin2016direct, huder2018electronic,yankowitz2012emergence, cao2018unconventional, yankowitz2019tuning, lu2019superconductors, sharpe2019emergent, seyler2019signatures, tran2019evidence, jin2019observation, wang2020correlated}, including the flat band \cite{bistritzer2011moire, cao2018correlated}, fractional Chern insulators \cite{chen2020tunable, abouelkomsan2020particle}, quantum Hall effect \cite{dean2013hofstadter,serlin2020intrinsic,bultinck2020mechanism,wu2020collective}, moir\'{e} ferroelectricity \cite{zheng2020unconventional,zheng2020unconventional}, exotic magnetism \cite{sharpe2019emergent,seo2019ferromagnetic,liu2019quantum,tschirhart2021imaging}, moir\'{e} excitons \cite{seyler2019signatures,alexeev2019resonantly,andersen2021excitons}, non-Abelian gauge potentials \cite{san2012non,ahn2019failure,park2019higher,nuckolls2020strongly}, and unconventional superconductivity \cite{cao2018correlated,cao2018unconventional,yankowitz2019tuning,lu2019superconductors}.
In parallel to these developments in solid state materials, examples of twisted-bilayer are also studied in other artificial quantum systems, such as in optics \cite{huang2016localization,wang2020localization,fu2020optical,mao2021magic,arkhipova2023observation}, acoustics \cite{gardezi2021simulating,zheng2022topological,jiang2022phononic,duan2023synthetic} and cold atoms  \cite{soltan2011multi,tarruell2012creating,jo2012ultracold,gall2021competing,meng2023atomic}.
In photonic lattices, the moir\'{e} pattern and its induced phenomena, such as the localization of light and flat band physics, have been studied in the photorefractive medium \cite{huang2016localization,wang2020localization,fu2020optical} and nanostructured photonic moir\'{e} superlattices \cite{mao2021magic}.

Ultracold atoms in optical lattices are another one of the most promising platform to
investigate the remarkable physical properties of twisted artificial structures \cite{soltan2011multi,tarruell2012creating,jo2012ultracold,gall2021competing,meng2023atomic}.
It thus has stimulated tremendous amount of efforts,  including theoretically predicting moir\'{e} physics in ultracold fermions \cite{salamon2020quantum,salamon2020simulating,luo2021spin,lee2022emulating} and bosons  \cite{gonzalez2019cold}. In particular,  experimental breakthrough observation of the spatial moir\'{e} pattern and momentum diffraction of a Bose-Einstein condensate (BEC) in twisted bilayer optical lattices has recently been achieved \cite{meng2023atomic}.
However, the conventional wisdom in past studies focuses on achieving the static twisted-bilayer lattices~\cite{meng2023atomic,salamon2020quantum,salamon2020simulating,luo2021spin,lee2022emulating,gonzalez2019cold}.
How to realize such twisted lattice structures in a dynamical way has so far eluded efforts.

\begin{figure}
	\centering
	\includegraphics[width=0.45\textwidth]{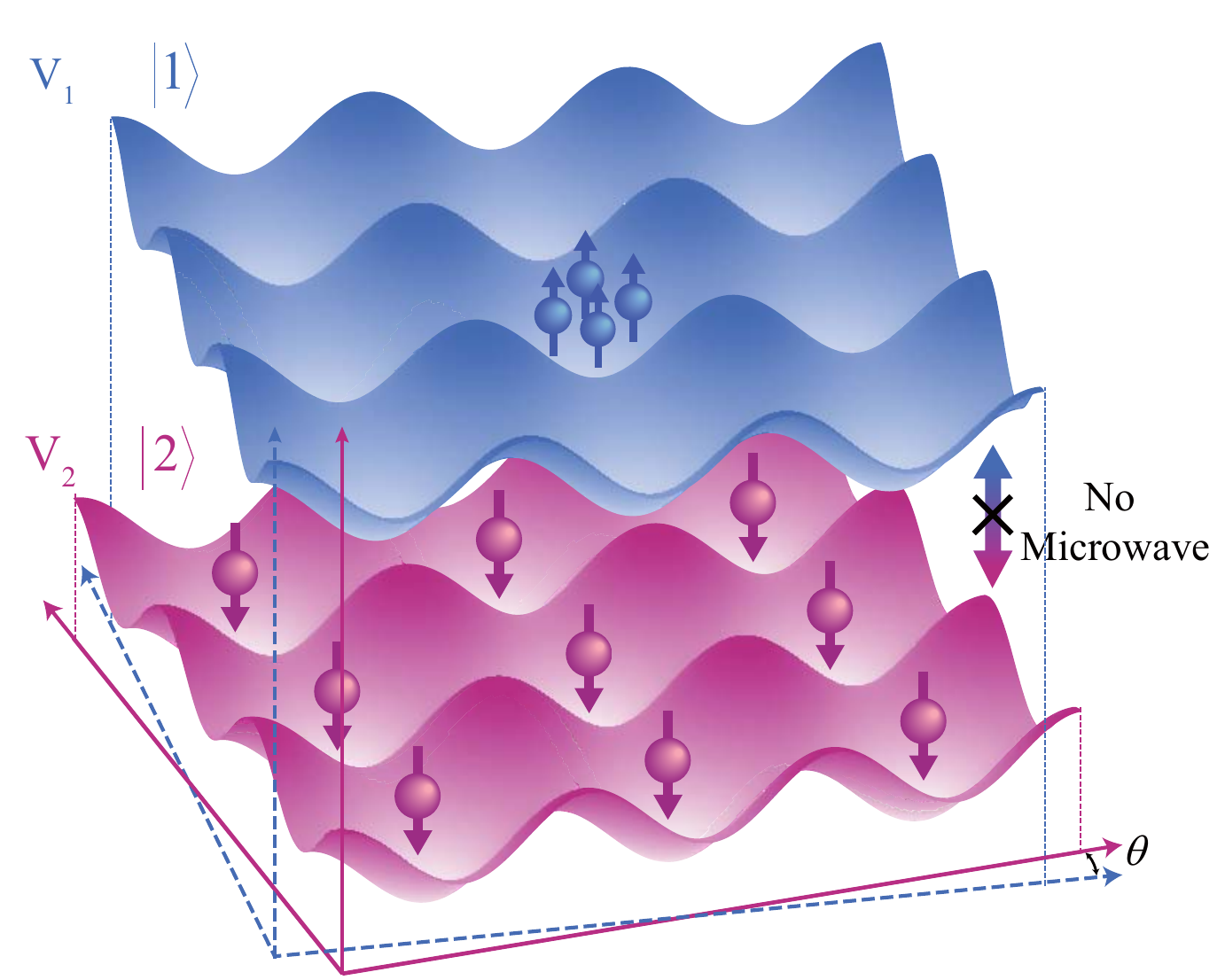}
	\caption{
		Schematic illustration of the system. A two-component BEC is considered here. Component 1 atoms are trapped in a tight harmonic trap and component 2 atoms are loaded in a square optical lattice confined by a very shallow harmonic trapping potential. After the system reaches equilibrium, the tight harmonic trap for component 1 atoms is tuned to be very shallow and a square optical lattice is also added. Distinguished from previous schemes via adding external single-particle type interlayer coupling \cite{meng2023atomic,luo2021spin,gonzalez2019cold}, such as adding microwave fields between two layers, our scheme only relies on the nonlinear effect from the intrinsic atomic interactions.
	}
	\label{fig::lattice}
\end{figure}

In this work, we propose a new scheme to dynamically achieve twisted-bilayer lattices that builds on the concept of nonlinearity. Our scheme relies on the  nonlinear interaction effect, which naturally emerges in the Gross-Pitaevskii equations of interacting bosonic ultracold atoms. We find that such nonlinear interaction effect amazingly results in a dynamical self-organized twisted-bilayer lattice. This new scheme frees out the usual requirement of adding external single-particle type coupling between different layers and only relies on the nonlinear effect from the intrinsic atomic interactions,  dramatically distinguished from previous schemes ~\cite{meng2023atomic,luo2021spin,gonzalez2019cold}. We show that both dynamical self-organized periodic (commensurable) and aperiodic (incommensurable) twisted-bilayer lattice structures can be realized through our proposed scheme. In particular, for the aperiodic case, we find the localization of the Gaussian BEC wave packet, resulting from the emergence of the flat bands in the nonlinearity-induced self-organized twisted-bilayer lattices. It indicates that our proposed nonlinear interaction effect can dynamically produce the tunable moir\'{e} lattices in BEC.

\section{Effective model}
The system considered here is schematically illustrated in Fig.\ref{fig::lattice}. A two-component BEC, for instance $^{87}$Rb atoms, is used,  where a spin-dependent harmonic potential is applied \cite{PhysRevAadd,PhysRevLettadd,pegahan2019spin}. Atoms in component 1 are trapped in a tight harmonic trap and the atoms in component 2 are loaded in a square optical lattice confined by a very shallow harmonic trapping potential. After the system reaches equilibrium, the tight harmonic trap for component 1 atoms is tuned to be very shallow and a square optical lattice is added. At the same time, the inter-species interaction is turned on. Such a system can be described by the following model Hamiltonian
\begin{eqnarray}
\mathbf{H}&=&\int \text{d}x\text{d}y \sum_{\alpha=1,2} \psi_\alpha^\dagger \left[ \frac{\mathbf{k}^2}{2m}+V_{\text{trap},\alpha}(x,y)+V_{\alpha}(x,y) \right] \psi_\alpha \notag \\
&+&\int \text{d}x\text{d}y(g_{11}\hat{n}_1^2+g_{22}\hat{n}_2^2+2g_{12}\hat{n}_1\hat{n}_2),	
\label{eq::GP H}
\end{eqnarray}
where $\psi_\alpha$ $(\alpha=1,2)$ denote the Bose field operators for two-component bosons, respectively. The corresponding density operators  are defined as $\hat{n}_\alpha=\psi^\dagger_\alpha\psi_\alpha$. $V_{\text{trap},\alpha}$ denotes the harmonic trap. The spin-dependent optical  lattices are  $V_1(x,y)={V}_1[\sin^2(\cos\frac{\theta}{2}k_Lx-\sin\frac{\theta}{2}k_Ly)+\sin^2(\sin\frac{\theta}{2}k_Lx+\cos\frac{\theta}{2}k_Ly)]$ and  $V_2(x,y)={V}_2[\sin^2(\cos\frac{\theta}{2}k_Lx+\sin\frac{\theta}{2}k_Ly)+\sin^2(-\sin\frac{\theta}{2}k_Lx+\cos\frac{\theta}{2}k_Ly)]$, respectively, with $\theta$ being the twist angle. $k_L=2\pi/\lambda$ is the wavevector
of the laser field with the corresponding lattice constant defined as $a_L=\pi/k_L$, and $\lambda$ is the wavelength of the laser. ${V}_\alpha$ captures the corresponding lattice depth.
$g_{11}$ and $g_{22}$ characterize the intra-species interaction strength for component 1 and  component 2 atoms, respectively. The inter-species interaction strength is captured by $g_{12}$.

From Eq. (1), the following
dimensionless energy functional can be constructed under the Gross-Pitaevskii (GP) mean-field theory
\begin{eqnarray}
\varepsilon&=&\int \text{d}x'\text{d}y' \sum_{\alpha=1,2}
		\psi_\alpha^* \left[ -\frac{\nabla^2}{2}+V'_{\text{trap},\alpha}(x',y')+V'_{\alpha}(x',y') \right] \notag \\
		&\times& \psi_\alpha +\int \text{d}x'\text{d}y'(g'_{11}|\psi_1|^4+g'_{22}|\psi_2|^4+2g'_{12}|\psi_1|^2|\psi_2|^2),\notag \\
\label{eq::GP E}
\end{eqnarray}
where dimensionless spatial coordinates are defined as $x'=x/\lambda$ and $y'=y/\lambda$, respectively. $V'_{\text{trap},\alpha}=V_{\text{trap},\alpha}/(\hbar^2/m\lambda^2)$ represents the harmonic trapping potential. $V'_{\alpha}=V_{\alpha}/(\hbar^2/m\lambda^2)$ labels the dimensionless lattice depth. $g'_{11(12,22)}=g_{11(12,22)}/(\hbar^2/m)$ are the corresponding dimensionless interaction strength.
The wavefunction is normalized as $\iint |\psi_{\alpha}|^2\text{d}x'\text{d}y'=N_{\alpha}$ with $N_\alpha$ being the particle number of the two-component atoms, respectively.
Based on Eq. (2), the dynamics of the system can be well described by the following coupled GP equations
\begin{subequations}\label{eq::GP}
	\begin{align}
			i\frac{\partial \psi_1}{\partial t'} = &\Big[ -\frac{1}{2}\nabla^2+V'_{\text{trap,1}}(x',y')+V'_1(x',y')+g'_{11}|\psi_1|^2 \nn \\
			&+g'_{12}|\psi_2|^2 \Big] \psi_1  \label{eq::GP1} \\
			i\frac{\partial \psi_2}{\partial t'} =& \Big[ -\frac{1}{2}\nabla^2+V'_{\text{trap,2}}(x',y')+V'_2(x',y')+g'_{12}|\psi_1|^2 \nn	\\
			&+g'_{22}|\psi_2|^2 \Big] \psi_2,		\label{eq::GP2}
		\end{align}
\end{subequations}
where $t'=t/(m\lambda^2/\hbar)$ denotes the  dimensionless evolution time.
\begin{figure}
	\centering
	\includegraphics[width=0.51\textwidth]{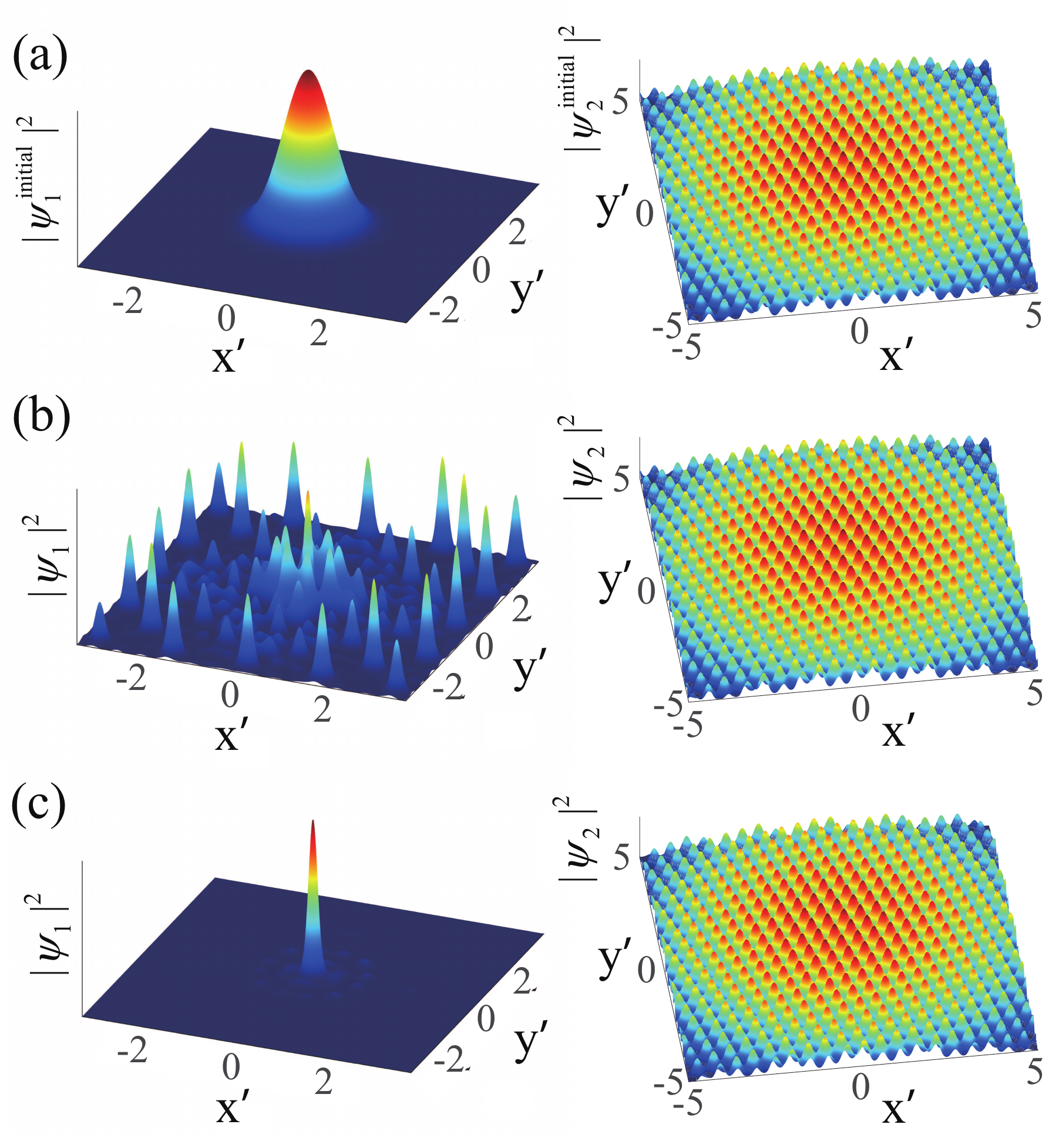}
	\caption{Time evolution of the two-component BEC in the nonlinearity-induced dynamical self-organized twisted-bilayer lattices. The left column of (b) and (c) show the distinct dynamics of the component 1 BEC with different twist angles $\theta$, starting from a Gaussian wave packet as the initial state shown in the left column of (a). In the presence of nonlinear term $w$ (defined in the main text), even without external single-particle type coupling between two layers, one of the interesting moir\'{e} physics, i.e., both the delocalized and localized behavior of the Gaussion wave packet of component 1 atoms, can be achieved when $\theta$ being chosen as a Pythagorean angle, $\theta= \arctan3/4$ in (b) and a non-Pythagorean angle $\theta = \pi/6$ in (c), respectively. The right column shows the time evolution of the component 2 BEC. Here we choose $g'_{11}=1.85\times 10^{-4}$, $g'_{22}=2\times 10^{-2}$, $g'_{12}= 2.1\times 10^{-3}$, $V'_1= 36 $, $V'_2= 333$, $N_1=1\times10^3$ and $N_2=5\times10^5$. In (b) and (c), the evolution time is chosen as $t'=10$.}
	\label{fig::psi}
\end{figure}

\section{Moir\'{e} localization induced by the  nonlinear interaction effect}
In the following, we will demonstrate that distinguished from previous studies \cite{meng2023atomic,luo2021spin,gonzalez2019cold}, even without
external single-particle type interlayer coupling,  the existence of nonlinear interaction term in GP equation, i.e., $w=g'_{11}|\psi_1|^2
+g'_{12}|\psi_2|^2$, can dynamically result in the twisted-bilayer lattices and  lead to one of the interesting moir\'{e} phenomena.
To show that, we monitor the dynamics of BEC for different twist angles $\theta$. Here the initial state for component 1 atoms is the ground state of the harmonic trap, which is chosen as a Gaussian wave packet (left part in Fig.~\ref{fig::psi}(a)). While for component 2 atoms, the initial state is chosen as the ground state (right part in Fig.\ref{fig::psi}(a)) determined by the effective Hamiltonian in the right hand-side of Eq.~\eqref{eq::GP2} with $g'_{12}=0$.

Interestingly, as shown in Fig.~\ref{fig::psi}, when tuning the twist angle $\theta$, in the presence of nonlinear term $w$, even without external single-particle type coupling between two layers, one of the interesting moir\'{e} physics, i.e., both the localized and delocalized behavior of component 1 atoms, can be achieved.
For instance, as shown in Fig.~\ref{fig::psi}(b) and (c), when $\theta$ is chosen as a Pythagorean angle, for instance, $\theta= \arctan3/4$ (Fig.~\ref{fig::psi}(b)), the Gaussion wave packet for component 1 atoms displays the delocalized behavior.
While considering $\theta$ being the non-Pythagorean twisted angle, for instance, $\theta = \pi/6$ (Fig.~\ref{fig::psi}(c)), the wave packet for component 1 atoms turns out to be localized.
Note that here we consider the case with large number imbalance between two components, i.e., $N_1 \ll N_2$ and lager intra-species interaction, i.e., $g'_{22} > g'_{12}$. The effect of inter-species interaction on the dynamics of component 2 atoms is highly suppressed and the density profile of component 2 atoms almost does not change during the evolution, as shown in the right column of Fig.~\ref{fig::psi}.

\begin{figure}
	\centering
	\includegraphics[width=0.52\textwidth]{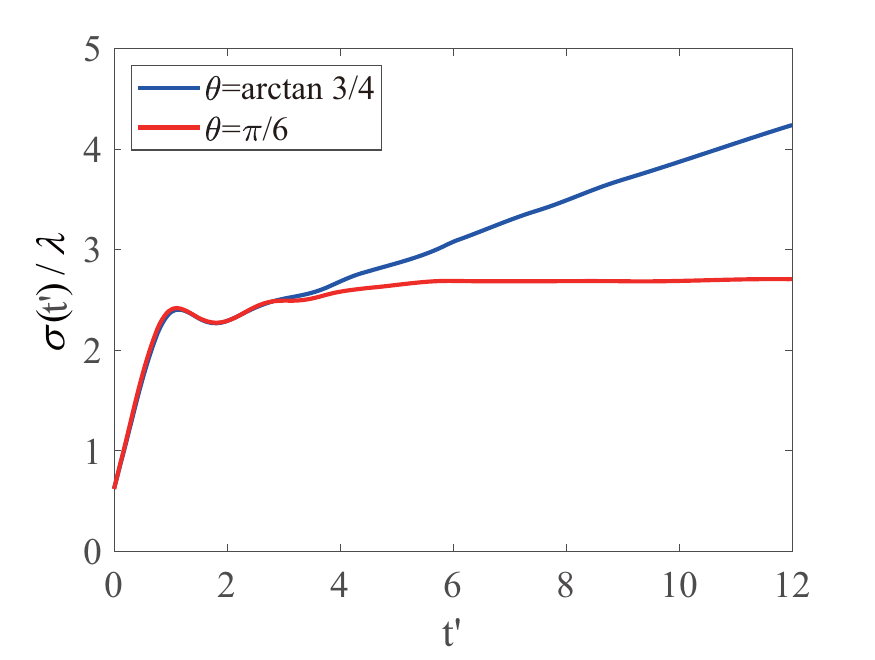}
	\caption{
		Mean square displacement of the component 1 BEC wave packet as the function of the evolution time.
		When the twist angle $\theta$ being a Pythagorean angle, i.e., $\theta= \arctan3/4$, $\sigma(t)$ tends to linearly
		increase with respect to the evolution time, i.e., $\sigma(t) \propto t$, indicating its delocalized behavior.
		While $\theta$ being a non-Pythagorean angle, i.e., $\theta = \pi/6$, $\sigma(t)$ reaches a stable value in a fairly short time, characterizing its localized behavior. Other parameters are chosen as the same in Fig.~\ref{fig::psi}.
	}
	\label{fig::MSD}
\end{figure}

To quantitatively distinguish between the  delocalized and localized behavior of the wave packet of component 1 atoms, we analyze the following mean square displacement (MSD)~\cite{JPSJ1988,PhysRevLettzhong2001}
\begin{equation}\label{key}
	\sigma(t)=\left[\frac{\iint \text{d}x\text{d}y (x^2+y^2)|\psi_1(t)|^2}{\iint \text{d}x\text{d}y|\psi_1(t)|^2}\right]^{\frac{1}{2}},
\end{equation}
which can characterize the width of the wave packet of component 1
atoms. The different dynamical behaviors of the wave packet can be
quantified as $\sigma(t)\propto t^{\kappa}$ with distinct dynamical index $\kappa$.  As shown in Fig.~\ref{fig::MSD},  when the twist angle $\theta$ is chosen as a Pythagorean angle, for instance, $\theta= \arctan3/4$, $\sigma(t)$ tends to linearly increase with respect to the evolution time, i.e., $\sigma(t) \propto t$.  The dynamical index is 1,  indicating
its delocalized behavior. While $\theta$ being a non-Pythagorean angle, for instance, $\theta = \pi/6$, $\sigma(t)$ will reach a stable value in a fairly short time, indicating the dynamical index $\kappa$ being 0. It thus characterizes the localized behavior of the wave packet.  Therefore, through tuning the twist angle $\theta$, one of the moir\'{e} physics, i.e., both the localized and delocalized behavior of the wave packet of component 1 atoms, can be achieved.

\section{Nonlinearity-induced dynamical self-organized twisted-bilayer lattices}
\begin{figure}
	\centering
	\includegraphics[width=0.51\textwidth]{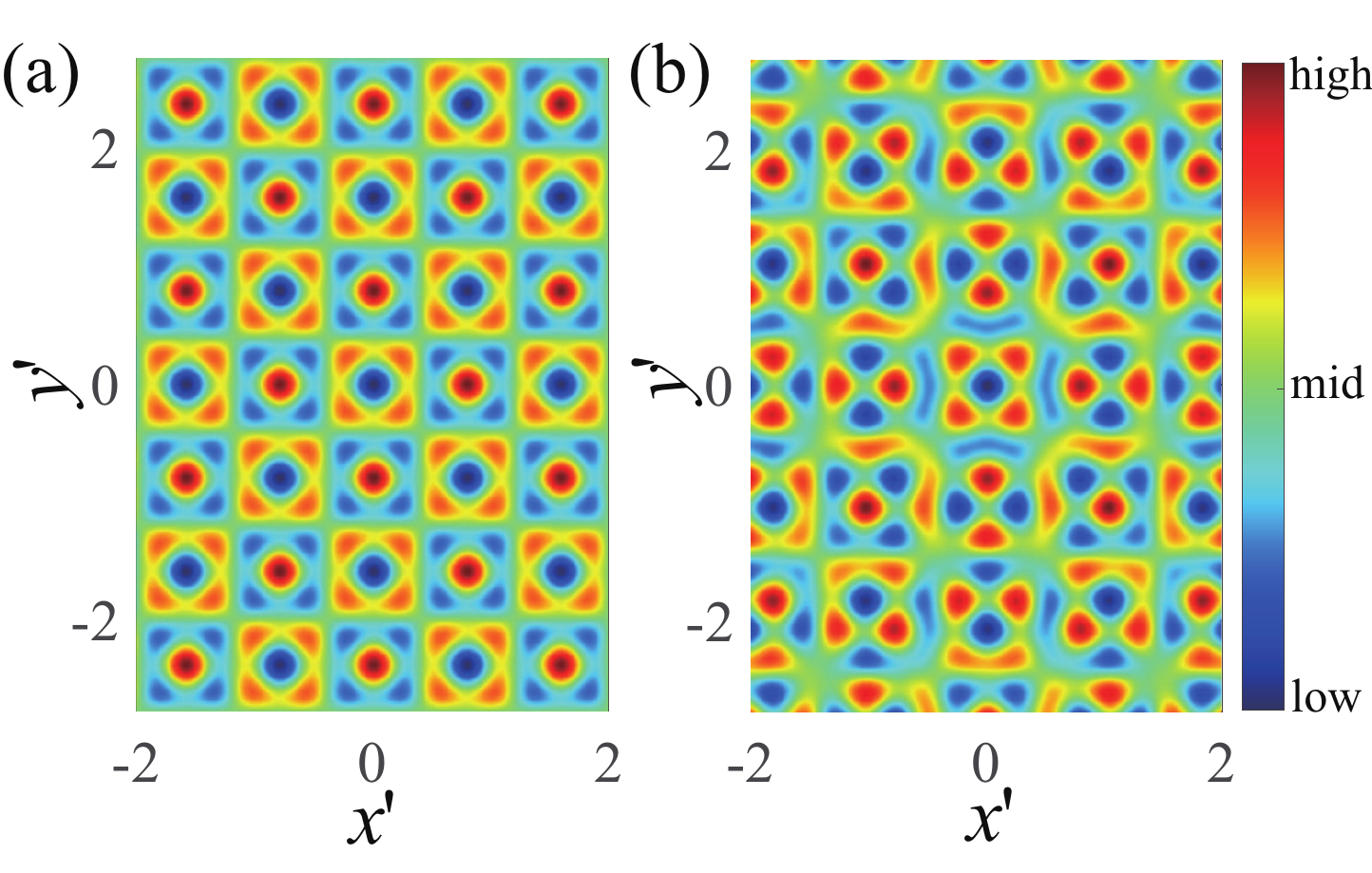}
	\caption{Nonlinearity-induced  self-organized twisted-bilayer lattice $V_{eff}$ for component 1 BEC. Both periodic (commensurable) and aperiodic (incommensurable) structures can be created in such a nonlinearity-induced twisted-bilayer lattice through adjusting the twist angle $\theta$ from the Pythagorean angle to the non-Pythagorean angle , where in (a) $\theta= \arctan3/4$ and in (b) $\theta = \pi/6$, respectively. Here, the evolution time is chosen as $t'=10$. Other parameters are chosen as the same in Fig.~\ref{fig::psi}.
	}
	\label{fig::Veff}
\end{figure}

In the following, we will explain the reason for the appearance of the moir\'{e} physics in our proposed system. Let us focus on Eq.~\eqref{eq::GP1} in the GP equation describing the two-component atomic BEC. In the presence of the nonlinear term $w$,  component 1 atoms suffer from a  dynamical self-organized effective potential $V_{eff}\equiv V'_{\text{trap,1}}+V'_{1}+w=V'_{\text{trap,1}}+V'_{1}+g'_{11}|\psi_1|^2+g'_{12}|\psi_2|^2$.
As shown in Fig.~\ref{fig::Veff}(a), when the twist angle $\theta$ is chosen as a Pythagorean
angle, for instance, $\theta =\arctan 3/4$, the effective potential $V_{eff}$ possess a spatially periodic structure. While considering $\theta$ being a non-Pythagorean twist angle, for instance,  $\theta=\pi/6$, as shown in Fig.~\ref{fig::Veff}(b), $V_{eff}$ behaves as a twisted-bilayer lattice with a spatial aperiodic structure. Therefore, distinguished from previous studies \cite{meng2023atomic,luo2021spin,gonzalez2019cold}, even without external single-particle type interlayer coupling, the nonlinear effect resulting from the intrinsic atomic interactions provides a new way to dynamically induce the twisted-bilayer lattices. Through tuning the twist angle $\theta$, we find that both periodic (commensurable) and aperiodic (incommensurable) moir\'{e} structures can be created in such a dynamical twisted-bilayer lattice.

To further understand the distinct dynamical behaviors of the wave packet of component 1 atoms, we rely on the analysis of the appearance of the flat band physics in our proposed nonlinearity-induced self-organized twisted-bilayer lattices. To do that, we calculate the band structure of $V_{eff}$ for both periodic (commensurable) and aperiodic (incommensurable) moir\'{e} structures.
In particular, when considering the aperiodic moir\'{e} lattice, the band structure can be calculated through approximating the non-Pythagorean twist angle by a Pythagorean one. For instance, here we use $\theta=\arctan 120/209$ to approximate $\theta= \pi/6$.
The band structure of $V_{eff}$ can thus be obtained by the plane-wave expansion method through introducing the Bloch basis $\psi_{n\vec{k}}=\sum_{\vec{G}} u_{n\vec{k},\vec{G}}|\vec{k}+\vec{G}\rangle$ with the Bloch vector $\vec{k}$ and the reciprocal lattice vector $\vec{G}$. Here $n$ labels the band index. Then, the energy spectra can be obtained through solving the eigen-problem via the following relation
\begin{equation}
	\begin{aligned}
		\frac{(\vec{k}+\vec{G})^2\lambda^2}{2}u_{n\vec{k},\vec{G}} + \sum_{\vec{G'}}\langle \vec{k}+\vec{G} | V_{eff}| \vec{k}+\vec{G'}\rangle u_{n\vec{k},\vec{G'}}& \\
		= E'_{n\vec{k}} u_{n\vec{k},\vec{G}},&
	\end{aligned}	
\end{equation}
where $E'_{n\vec{k}}=E_{n\vec{k}}/(\hbar^2/m\lambda^2)$ is the dimensionless energy spectra.

\begin{figure}
	\centering
	\includegraphics[width=0.505\textwidth]{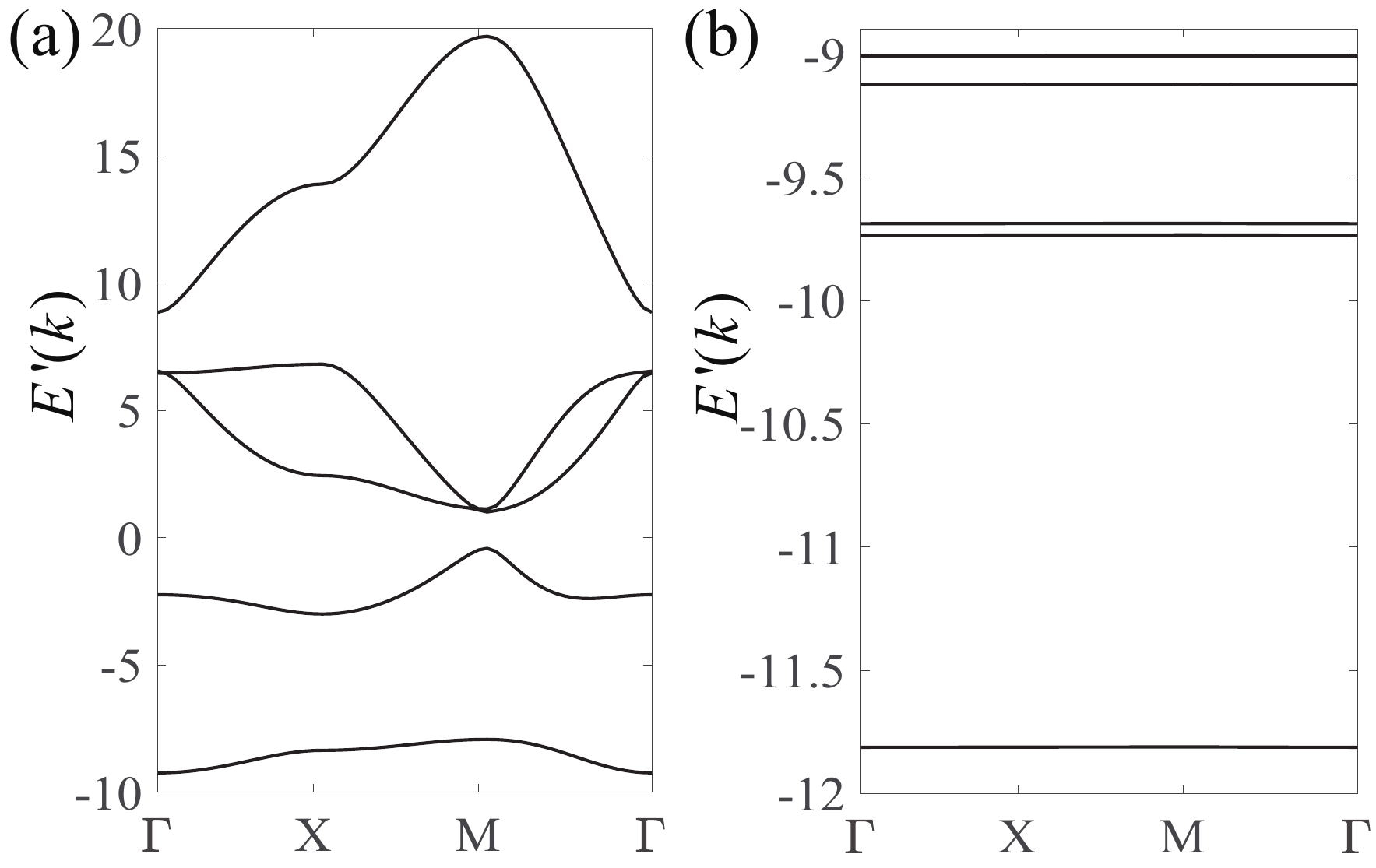}
	\caption{
		The band structure of the nonlinearity-induced self-organized twisted-bilayer lattice. (a) and (b) show the corresponding band structure (the lowest five bands) of the moir\'{e} lattices $V_{eff}$ demonstrated in Fig.~\ref{fig::Veff} (a) and (b), respectively.
		It is shown that the flat bands emerge when changing the twist angle $\theta$ from the Pythagorean angle (in (a)) to the non-Pythagorean angle (in (b)),  accompanying with
the lattice structure changing from the periodic case to the aperiodic case. Other parameters are chosen as the same in Fig.~\ref{fig::Veff}.
	}
	\label{fig::band}
\end{figure}

As shown in Fig.~\ref{fig::band}, the band structures of the corresponding
twisted-bilayer lattices $V_{eff}$  demonstrated in Fig.~\ref{fig::Veff} are obtained. We find that when changing the twist angle from a Pythagorean angle to a non-Pythagorean angle, more dispersive lower bands become extremely flat, accompanying with the lattice structure changing from the periodic (commensurable) case to the aperiodic (incommensurable) one. In particular, the appearance of the lowest flat band will lead to the localization of the wave packet of component 1 atoms, since such a BEC wave packet will mainly concentrate at the lowest band after a long enough time evolution. Furthermore, it is known that the flat band can support quasi-nondiffracting localized modes. Therefore, the wave packet of component 1 atoms loaded in such a  nonlinearity induced self-organized aperiodic (incommensurable) twisted-bilayer lattice will tend to be localized.

\section{Conclusion}
In summary, we have demonstrated a new approach to
dynamically achieve twisted-bilayer lattices in an atomic BEC.
To emphasize a remarkable difference, our scheme
does not require external single-particle type
interlayer coupling. Instead, we only require the nonlinear effect
arising from the intrinsic atomic interactions.
We further show that both periodic and aperiodic  moire lattices, i.e.,  nonlinearity-induced dynamical self-organized twisted-bilayer lattices, can be designed through tuning the twist angle. One of the interesting phenomena in twisted-bilayer lattices, i.e., the flat-band physics, is also shown through investigating the dynamics of the wave packet of BEC. Such nonlinearity-induced approach is rather generic to other quantum systems, such as in photonics and electrical circuits, than restricted to the setup considered in this work, since the nonlinear effects, such as  Kerr nonlinearity, is common in these systems. Our results reveal a profound connection between
the nonlinearity and twistronics.
\section{Acknowledgment}
We acknowledge helpful discussions with L. Chen. This work is supported by the National Key R$\&$D Program of China (2021YFA1401700), NSFC (Grants No. 12074305, 12147137, 11774282), the Fundamental Research Funds for the Central Universities (Grant No. xtr052023002), the Shaanxi Fundamental Science Research Project for Mathematics and Physics (Grant No. 23JSZ003) and  Xiaomi Young Scholar Program. We also thank the HPC platform of Xi'an Jiaotong University, where our numerical calculations was performed.

\bibliographystyle{apsrev}
\bibliography{GP-ref}

\end{document}